\documentclass[aps,prd,superscriptaddress,onecolumn,showpacs,showkeys]{revtex4}
\usepackage{amssymb,amsmath,epsfig}

\usepackage{eurosym}
\usepackage{amsfonts}
\usepackage{array}
\usepackage{amsthm}
\usepackage{bm}
\usepackage{palatino}
\usepackage{mathpazo}
\usepackage{amssymb}
\usepackage{eurosym}
\usepackage{amsmath}
\usepackage{epsfig}
\usepackage{graphics}
\usepackage{color}
\usepackage{graphicx}
\usepackage{hyperref}
\usepackage{changes}

\begin{document}

\title{Deflection angle of photon from magnetized black hole and effect of nonlinear electrodynamics}

\author{Wajiha Javed} \email{wajiha.javed@ue.edu.pk;
wajihajaved84@yahoo.com}
\affiliation{Department of Mathematics, University of Education,\\
Township, Lahore-54590, Pakistan.}

\author{Jameela Abbas}
\email{jameelaabbas30@gmail.com}
\affiliation{Department of Mathematics, University of Education,\\
Township, Lahore-54590, Pakistan.}

\author{Ali \"{O}vg\"{u}n}
\email{ali.ovgun@pucv.cl}
\homepage[]{https://www.aovgun.com}
\affiliation{Instituto de F\'{\i}sica, Pontificia Universidad Cat\'olica de Valpara\'{\i}%
so, Casilla 4950, Valpara\'{\i}so, Chile}
\affiliation{Physics Department, Faculty of Arts and Sciences, Eastern Mediterranean
University, Famagusta, North Cyprus, via Mersin 10, Turkey}

\begin{abstract}
In this paper, we analyze deflection angle of photon from magnetized black hole within non-linear electrodynamics with parameter $\beta$.  In doing so, we find the corresponding optical spacetime metric
and then we calculate the Gaussian optical curvature. Using the Gauss-Bonnet theorem, we obtain the deflection angle of photon from magnetized black hole in weak field limits and show the effect of non-linear electrodynamics on weak gravitational lensing.  We also analyzed that our results reduces into Maxwell's electrodynamics and Reissner-Nordstr\"{o}m (RN) solution with the reduction
of parameters. Moreover, we also investigate the graphical behavior of deflection angle
w.r.t correction parameter, black hole charge and impact parameter.
\end{abstract}
\date{\today}

\maketitle
\section{Introduction}

In 1916, Einstein cleverly predicted the existence of gravitational waves and gravitational lensing as part of the theory of general relativity \cite{Einstein:1956zz}. In 2015, the gravitational waves were detected by LIGO \cite{Abbott:2016blz}, which shows that the theoretical predictions are well fitted with experimental observations. Hence, this new detection of gravitational waves marks not only a culmination of a decades-long search, but also the beginning of a new way to look at the universe. After detection of gravitational waves by LIGO, there is renewed interest in the topic of gravitational lensing\cite{Li:2018prc}. Gravitational lensing first proposed by Soldner in 1801 in context of Newtonian theory \cite{soldner}.  Many useful results for cosmology have come out of using this property of matter and light. Then using data taken during a solar eclipse in 1919, Eddington measured a value close to that of the GR prediction \cite{Dyson:1920cwa,VallsGabaud:2012xz}. Then gravitational lensing has been worked in various space-times using different methods \cite{Bartelmann:1999yn,Bartelmann:2010fz,Keeton:1997by,Eiroa:2002mk,Mao:1991nt,Bozza:2002zj,Gallo:2011mv,Crisnejo:2017jmx,Sharif:2015qfa}. 

Moreover, over the years, there have been many studies linking gravitational lensing with the Gauss-Bonnet theorem (GBT) after Gibbons and Werner (GW) elegantly showed that the possible way of calculation the deflection angle using the GBT for asymptotically flat static black holes \cite{Gibbons:2008rj}:  
\begin{equation}\notag
\alpha=-\int \int_{S_\infty} \mathcal{K}  \mathrm{d}\sigma,
\end{equation} 
Here $\mathcal{K}$ and $\mathrm{d}\sigma$ are the Gaussian curvature and surface element of optical metric.

Afterwards, Werner extended this method for stationary black holes \cite{Werner:2012rc}. Next, Ishihara et al. \cite{Ishihara:2016vdc} showed that it is possible to find deflection angle for the finite-distances (large impact parameter) because the GW only found the deflection angle using the optical Fermat geometry of the black hole's spacetime in weak field limits and for the observers at asymptotically flat region. Recently, Crisnejo and Gallo have studied the deflection of light in a plasma medium \cite{Crisnejo:2018uyn}.  Since then, there is a continuously growing interest to the
weak gravitational lensing via the method of GW and GBT of black holes, wormholes or cosmic strings \cite{Jusufi:2017lsl,Sakalli:2017ewb,Jusufi:2017mav,Ono:2017pie,Jusufi:2017vta,Ovgun:2018prw,Jusufi:2017hed,Arakida:2017hrm,Ono:2018ybw,Jusufi:2017uhh,Ovgun:2018xys,Jusufi:2018jof,Ovgun:2018fnk,Ovgun:2018ran,Ovgun:2018oxk,Ovgun:2018fte,Ovgun:2018tua,Ono:2018jrv,Ovgun:2019wej,Ovgun:2019xxx,Ovgun:2019ygw}.

The main aim of this paper is to investigate the effect of the nonlinear electrodynamics (magnetized) charge on the deflection angle where we use the GBT in which the deflection of light become a global effect \cite{Kruglov:2017fuj}. Because we only focus  the nonsingular region outside of a photon rays.

Gravitational singularities are mainly considered within general relativity, where density apparently becomes infinite at the center of a black hole, and within astrophysics and cosmology as the earliest state of the universe during the Big Bang. In general theory of gravity, spacetime singularities raise a number of problems, both mathematical and physical \cite{Kruglov:2018lct,Kruglov:2017xmb,Kruglov:2017mpj,Kruglov:2017fck,Kruglov:2016uzf,Kruglov:2016ymq,Kruglov:2016ezw,Novello:1999pg,Novello:2001fv}. Using the nonlinear electrodynamics it is possible to remove these singularities by constructing a regular black hole solution   \cite{Bronnikov:2000vy,AyonBeato:1998ub,AyonBeato:1999ec,Rodrigues:2018bdc,Hayward:2005gi}. Recently, Kruglov has proposed a new
model of nonlinear electrodynamics with two parameters $\beta$ and $\gamma$ where the the specific range of magnetic field, the causality and
unitary principles are satisfied \cite{Kruglov:2017fuj}. Moreover, it is shown that  there is no singularity of the
electric field strength at the origin for the point-like particles and it has a magnetic charge. Moreover, AN Aliev et al. showed the effect of the magnetic field on the
black hole space time \cite{Aliev:1989wx,Aliev:2006yk,Aliev:2006tt,Aliev:2004ec,Aliev:2002nw}.

This paper is organized as follows. In section 2, we briefly review the solution of magnetized black hole and then we calculate its optical geometry and the optical curvature. In section 3, deflection angle of photon using the GBT is studied in the case of magnetized black hole. In section 4, we analysis the deflection angle in details using the graphical analysis. And we concludes in section 4 with a discussion regarding the results obtained from the
present work.

\section{Weak Gravitational Lensing and Magnetized black hole}

The action for the Einstein with a nonlinear electrodynamics (NLED)  is given as \cite{Kruglov:2017fuj}
\begin{equation}
J=\int d^{4}x\sqrt{-g}(\frac{1}{2k^{2}}R+\mathcal{L}),
\end{equation}
where $k^{2}=8\pi G\equiv M^{-2}_{pl}$, $M_{pl}$ is for reduced planck mass, $G$ is Newtonian constant and $R$ is Ricci scalar. From above equation we derived the Einstein equations as
\begin{equation}
R_{ab}-\frac{1}{2}g_{ab}R=-k^{2}T_{ab}.
\end{equation}
Now, we derive the equation of motion for electromagnetic fields by varying $(1)$
\begin{equation}
\partial_{a} (\sqrt{-g}(F^{ab} \mathcal{L}_{F}+\bar{F}^{ab}\mathcal{L}_{G}))=0.
\end{equation}
Now, we analyze the static magnetic black hole solution by using the Einstein field equation and equation
of motion for electromagnetic field with above equations. In the case of pure magnetic field, Bronnikov showed\cite{Bronnikov:2000vy} that when spherical symmetry holds, the invariant is $\mathcal{F}=q^{2}/(2r^{2})$, where $q$ is a magnetic charge. In this case, the line element of the static and spherical symmetric
space-time is
\begin{equation}
d s^{2}=-f(r)dt^{2}+\frac{1}{f(r)}dr^{2}+r^{2}(d\theta^{2}+\sin^{2}\theta d\varphi^{2}),
\end{equation}
by assuming both the source and observer are in the equatorial plane likewise trajectory of the null photon is also on
the same plane with$(\theta=\frac{\pi}{2})$, we obtain the optical metric as follows
\begin{equation}
d s^{2}=-f(r)dt^{2}+\frac{1}{f(r)}dr^{2}+r^{2}d\varphi^{2}.
\end{equation}
For moving photon in the equatorial plane and null geodesics, $d s^{2}=0$, we get
\begin{equation}
dt^{2}=\frac{1}{f(r)^{2}}dr^{2}+\frac{r^{2}}{f(r)}d\varphi^{2}.
\end{equation}
Subsequently, we make the transformation into new coordinate $u$, the metric function $\zeta(u)$ as
\begin{equation}
du=\frac{dr}{f(r)} , \zeta=\frac{r}{\sqrt{f(r)}}.
\end{equation}
Then the optical metric tensor $\bar{g}_{ab}$ is as follows
\begin{equation}
dt^{2}=\bar{g}_{ab}dx^{a}dx^{b}=du^{2}+\zeta^{2}d\varphi^{2}.
\end{equation}
It is to be noted that $(a,b)\rightarrow (r,\varphi)$ and determinant is $det\bar{g}_{ab}=\frac{r^{2}}{f(r)^{3}}$. Now using $Eq.(8)$,
the non-zero Christopher symbols are $\Gamma^{r}_{rr}=-\frac{f'(r)}{f(r)}$, $\Gamma^{r}_{\varphi\varphi}=\frac{r(rf'(r)-2f(r))}{2}$, $\Gamma^{\varphi}_{r\varphi}=\frac{-rf'(r)+2f(r)}{2rf(r)}=\Gamma^{\varphi}_{\varphi r}$. Hence, we can find the Gaussian optical curvature $\mathcal{K}$ \cite{Gibbons:2008rj} as follows
\begin{equation}
\mathcal{K}=-\frac{R_{r\varphi r\varphi}}{det\bar{g}_{r\varphi}}=-\frac{1}{\zeta}\frac{d^{2}\zeta}{du^{2}}.
\end{equation}
Now, we rewrite Gaussian optical curvature in terms of Schwarzschild radial coordinate $r$\cite{B22}:
\begin{equation}
\mathcal{K}=-\frac{1}{\zeta}\left[\frac{dr}{du}\frac{d}{dr}(\frac{dr}{du})\frac{d\zeta}{dr}+(\frac{dr}{du})^{2}\frac{d^{2}\zeta}{dr^{2}}\right].
\end{equation}
By applying Eq.(10) into our metric (6) we obtain the Gaussian optical curvature of photon from magnetized black hole, which yields that
\begin{equation}
\mathcal{K}=-\frac{\sqrt{f(r)}}{r}\left[\frac{r(f'(r)^{2}-2f(r)f''(r))}{4\sqrt{f(r)}}\right],
\end{equation}
where the function $f(r)$ is \cite{Kruglov:2017fuj}
\begin{equation}
f(r)=1-\frac{2Gm}{r}+\frac{Gq^{2}}{r^{2}}-\frac{\beta Gq^{4}}{5r^{6}}+\mathcal{O}(r^{-10}).
\end{equation}
After substituting the value of $f(r)$, we get the value of optical curvature up to leading orders
\begin{equation}
\mathcal{K}\simeq-\frac{2 Gm}{r^{3}}+\frac{3Gq^{2}}{r^{4}}-\frac{21\beta G q^{4}}{5r^{8}}+O \left( {G}^{2},m^2 \right) .
\end{equation}
\section{Deflection angle of photon and Gauss-Bonnet theorem}
Now, we use the Gauss-Bonnet theorem to derive the deflection angle of photon for magnetized black hole. We apply the Gauss-Bonnet theorem to the region $\mathcal{M}_{R}$, stated as \cite{Gibbons:2008rj}
\begin{equation}
\int\int_{\mathcal{M}_{R}}\mathcal{K}dS+\oint_{\partial\mathcal{M}_{R}}kdt+\sum_{j}\theta_{j}=2\pi\mathcal{X}(\mathcal{M}_{R}).
\end{equation}
Here $\mathcal{K}$ is for Gaussian curvature and the geodesic curvature is $k$, given as $k=\bar{g}(\nabla_{\dot{\alpha}}\dot{\alpha},\ddot{\alpha})$ in such a way that $\bar{g}(\dot{\alpha},\dot{\alpha})=1$, where $\theta_{j}$ is the representation for exterior angle at the $j^{th}$ vertex and $\ddot{\alpha}$ is unit acceleration vector. The jump angles become $\pi/2$ as $R\rightarrow\infty$ and we get $\theta_{O}+\theta_{S}\rightarrow\pi$. The Euler characteristic is $\mathcal{X}(\mathcal{M}_{R})=1$, as $\mathcal{M}_{R}$ is non singular. Therefore we get
\begin{equation}
\int\int_{\mathcal{M}_{R}}\mathcal{K}dS+\oint_{\partial\mathcal{M}_{R}}kdt+\theta_{j}=2\pi\mathcal{X}(\mathcal{M}_{R}).
\end{equation}
Where $\theta_{j}=\pi$ shows the total jump angle and $\alpha_{\bar{g}}$ is a geodesic; since the Euler characteristic number $\mathcal{X}$ is $1$. As $R\rightarrow\infty$ the remaining part yield that $k(D_{R})=\mid\nabla_{\dot{D}_{R}}\dot{D}_{R}\mid$. The radial component of the geodesic curvature is given by \begin{equation}
(\nabla_{\dot{D}_{R}}\dot{D}_{R})^{r}=\dot{D}^{\varphi}_{R} \partial_{\varphi}\dot{D}^{r}_{R}+\Gamma^{r}_{\varphi\varphi}(\dot{D}^{\varphi}_{R})^{2}.
\end{equation}
At very large $R$, $D_{R}:=r(\varphi)=R=const$. Therefore, the first term of equation $(17)$ vanish and $(\dot{D}^{\varphi}_{R})^{2}=\frac{1}{\zeta}$. Recalling $\Gamma^{r}_{\varphi\varphi}=\frac{r(r f'(r)-2f(r))}{2}$, we get
\begin{equation}
(\nabla_{\dot{D}^{r}_{R}}\dot{D}^{r}_{R})^{r}\rightarrow\frac{-1}{R},
\end{equation}
and it shows that the geodesic curvature is independent of topological defects, $k(D_{R})\rightarrow R^{-1}$. But from the optical metric $(8)$, we can say that $dt=Rd\varphi$. Therefore we find that
\begin{equation}
k(D_{R})dt=\frac{1}{R}Rd\varphi.
\end{equation}
Taking into account the above results, we obtain
\begin{equation}
\int\int_{\mathcal{M}_{R}}\mathcal{K}dS+\oint_{\partial\mathcal{M}_{R}} kdt =^{R \rightarrow\infty }\int\int_{S_{\infty}}\mathcal{K}dS+\int^{\pi+\Theta}_{0}d\varphi.
\end{equation}
In the weak deflection limit, by assuming that at the zeroth order the light ray is given by $r(t)=b/\sin\varphi$. Thus by using $(14)$ and $(20)$, the deflection angle becomes \cite{Gibbons:2008rj}
\begin{equation}
\Theta=-\int^{\pi}_{0}\int^{\infty}_{b/\sin\varphi}\mathcal{K}\sqrt{det\bar{g}}dud\varphi,
\end{equation}
where,
\begin{equation}
\sqrt{det\bar{g}}du=r dr(1-\frac{6mG}{r}+\frac{3Gq^{2}}{r^{2}}-\frac{3q^{4}G\beta}{5r^{6}}).
\end{equation}
After substituting the leading order term of Gaussian curvature $(14)$ into equation$(21)$, the deflection angle up to second order term is calculated as follows:
\begin{equation}
\Theta\thickapprox \frac{4mG}{b}-\frac{3\pi Gq^{2}}{4b^{2}}+\frac{7G\pi \beta q^{4}}{32b^{6}}.
\end{equation}
\section{Graphical Analysis}
This section is devoted to discuss the graphically behavior of deflection angle $\Theta$. We also demonstrate the physical significance of these graphs to analyze the impact of correction parameter $\beta$, BH charge $q$, and impact parameter $b$ on deflection angle by examining the stability and instability of BH.
\subsection{Deflection angle with Impact parameter $b$}
This subsection gives the analysis of deflection angle $\Theta$ with impact parameter $b$ for different values of correction parameter $\beta$ and BH charge $q$ in geometrized unit $8\pi G = 1$. 

\begin{itemize}
\item  \textbf{Figure 1} shows the behavior of $\Theta$ with
$b$ for varying $q$ and for fixed value for correction parameter.
\begin{enumerate}
\item In figure (i), we examined that the deflection angle exponentially decreases for small variation of $q$.
\item In figure (ii), we analyzed that the deflection angle gradually decreasing and then eventually goes to infinity for large variation of $q$ which
is the unstable state of magnetized BH.\\
Therefore, we conclude that for small values of $q$ the magnetized BH is stable but as $q$ increases it shows the unstable behavior of magnetized BH. 
\end{enumerate}
In \textbf{Figure 2}, indicates the behavior of deflection angle with impact parameter by varying correction parameter $\beta$.
\begin{enumerate}
\item In figure (i), we noticed that the deflection angle decreasing constantly for small values of $\beta$ but in \item In figure (ii), as $\beta$ increases the deflection angle gradually decreasing and then goes to infinity.
\end{enumerate}    
\end{itemize}
\begin{center}
\epsfig{file=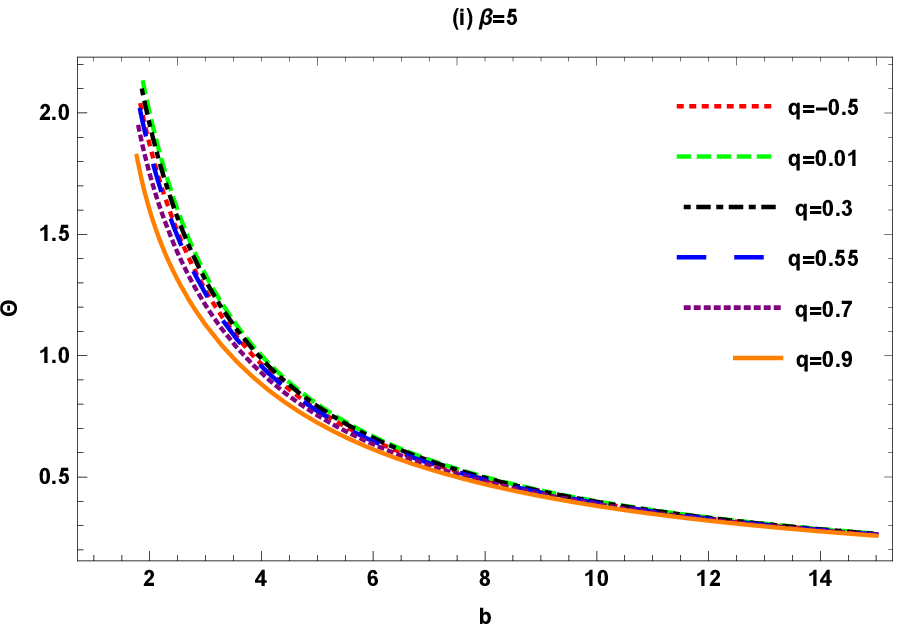,width=0.45\linewidth}\epsfig{file=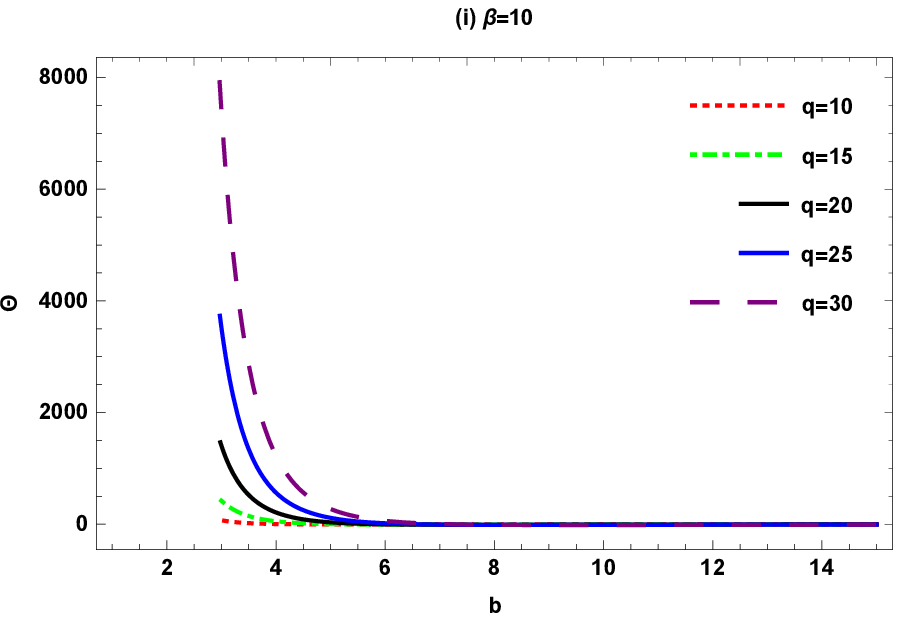,width=0.45\linewidth}\\
{Figure 1: Relation between $\Theta$ and impact parameter $b$}.\\
\end{center}
\begin{center}
\epsfig{file=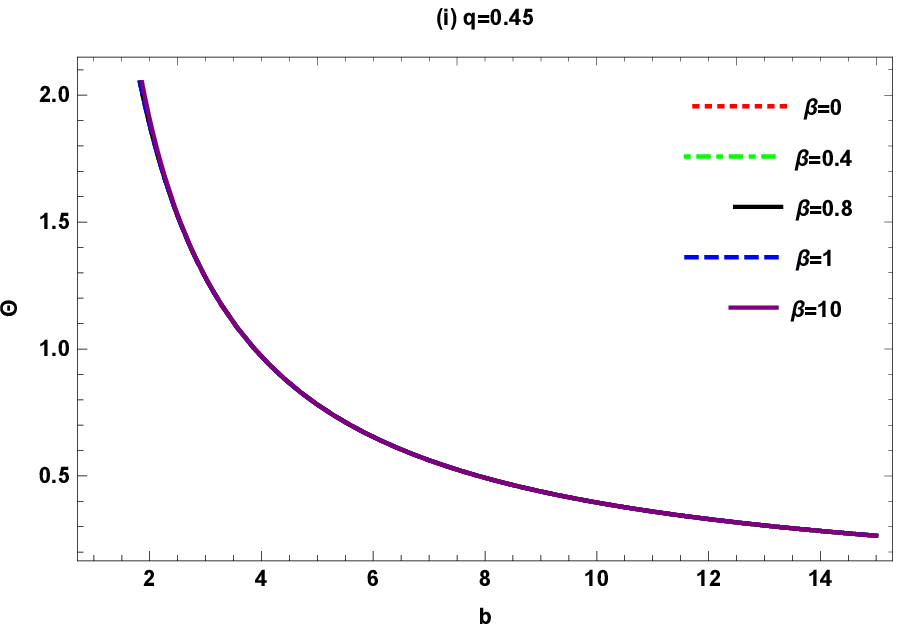,width=0.5\linewidth}\epsfig{file=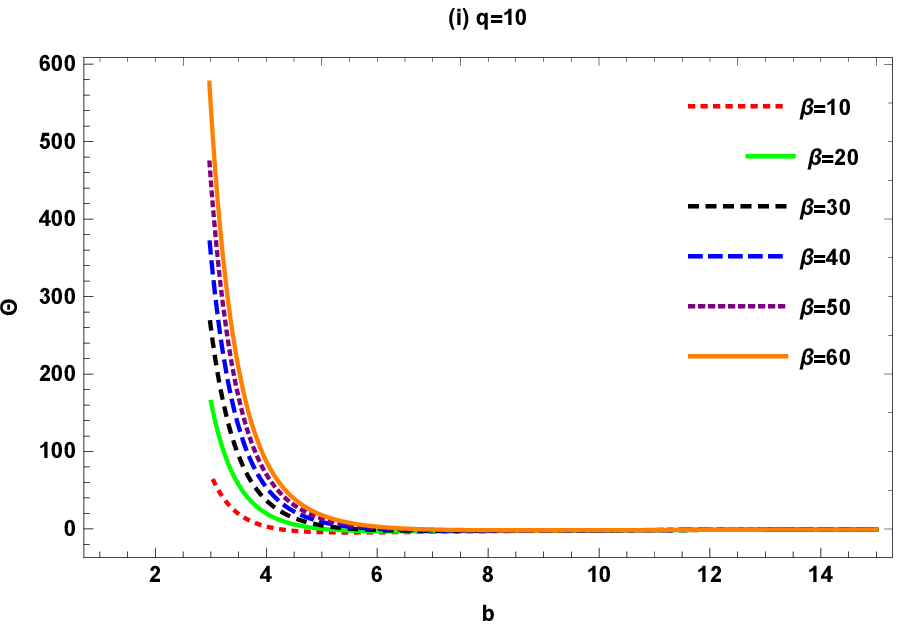,width=0.5\linewidth}\\
{Figure 2: Relation between $\Theta$ and impact parameter $b$}.
\end{center}
\begin{center}
\epsfig{file=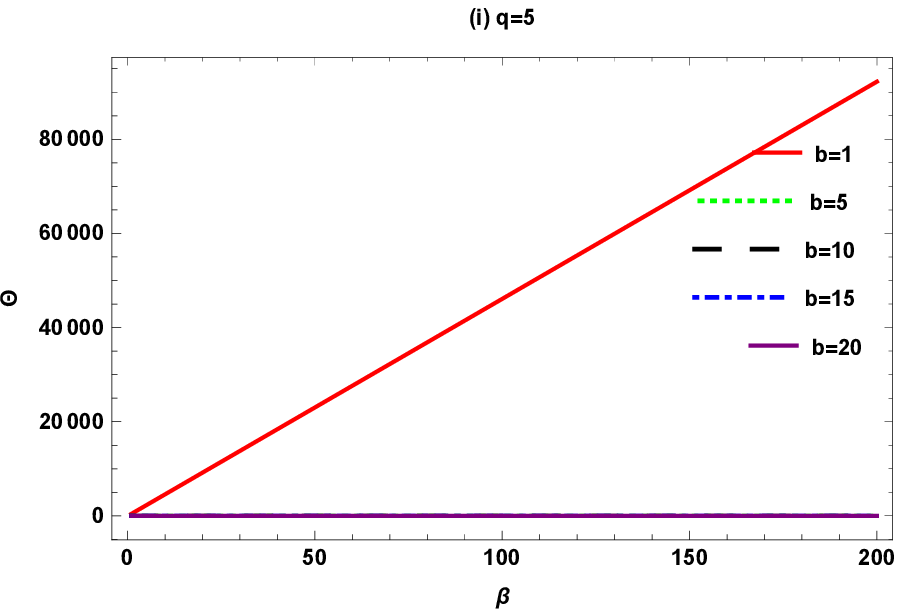,width=0.5\linewidth}\epsfig{file=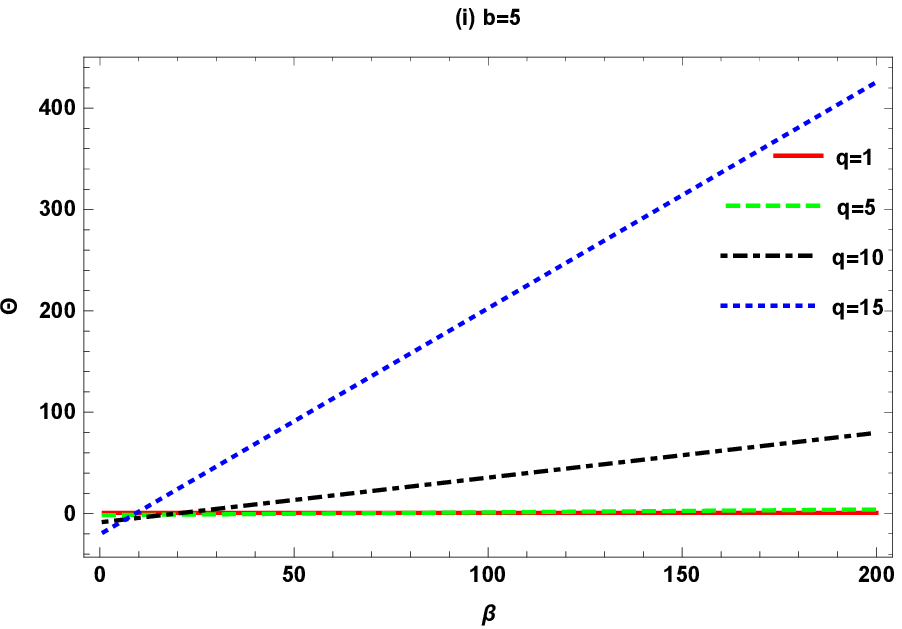,width=0.5\linewidth}
{Figure 3: Relation between $\Theta$ and correction parameter $\beta$}.
\end{center}
\begin{center}
\epsfig{file=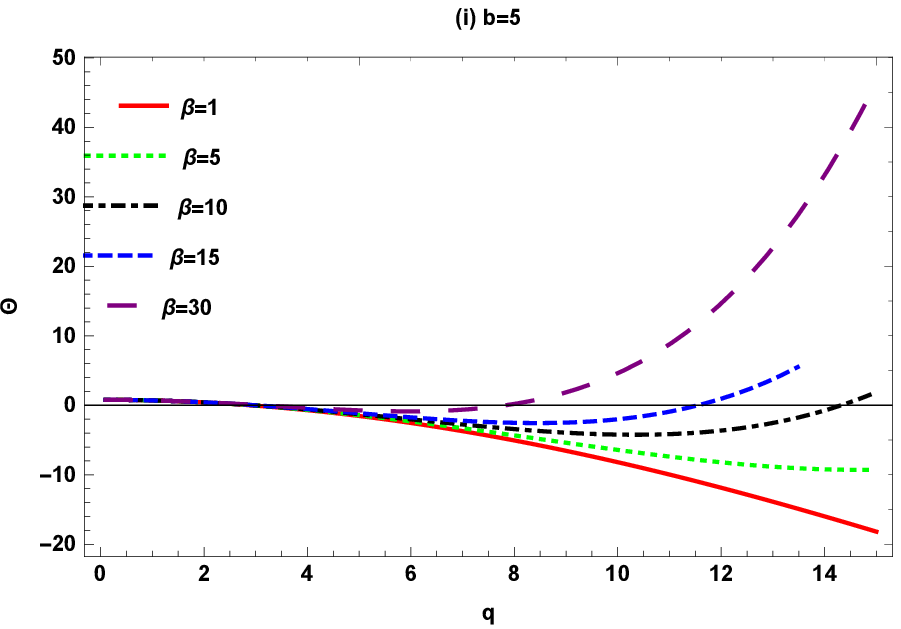,width=0.5\linewidth}\epsfig{file=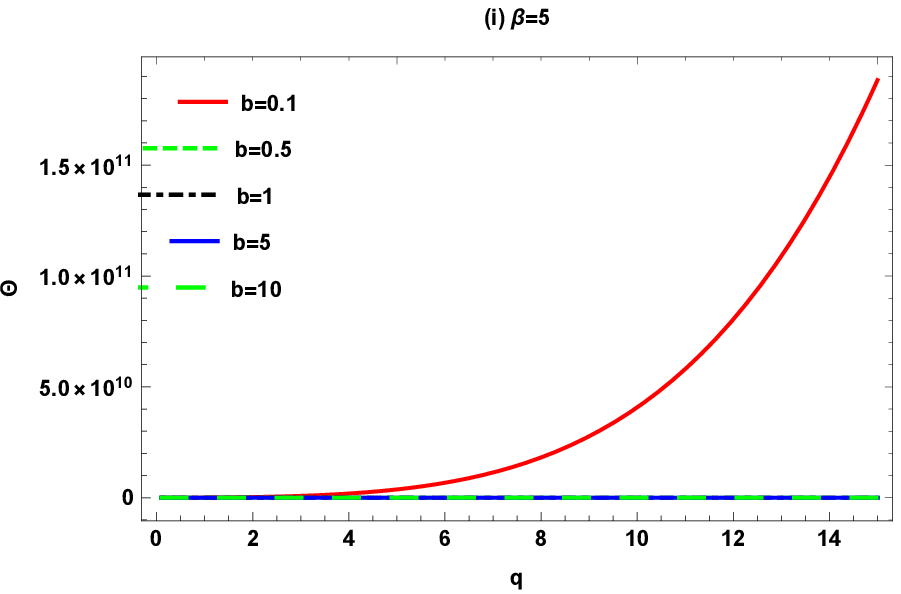,width=0.5\linewidth}
{Figure 4: Relation between $\Theta$ and BH charge $q$}.
\end{center}
\begin{itemize}
\item  \textbf{Figure 3} shows the behavior of $\Theta$ with correction parameter
$\beta$.
\begin{enumerate}
\item In figure (i), shows the behavior of $\Theta$ with $\beta$, for varying $b$ and fixed $q$. This shows that when $b<0$ it gives the uniform negative behavior but for $b>0$ it behave as a positive slope. We also conclude that for $4\leq b<15$ the behavior is negative slope but for $b\geq15$ it gives positive slope.
\item In figure (ii), indicate the behavior of $\Theta$ with $\beta$, for the variation of $q$ and fixed $b$. We examined that the deflection angle is positively increasing with increase of $q$ but for $5\leq q \leq 10$ the behavior is negative and then it becomes positive.\\
\end{enumerate}
In \textbf{Figure 4}, represent the behavior of deflection angle with BH charge $q$.
\begin{enumerate}
\item In figure (i), represents the behavior of $\Theta$ with $q$, by varying $\beta$ and fixed $b$. We analyzed that the deflection angle initially decreases but as $\beta$ increases the deflection angle firstly decreases and then increases. We also observe that for $\beta>20$ the behavior is increasing.
\item In figure (ii), shows the behavior of $\Theta$ with $q$, by varying $b$ and fixed $\beta$. We observed that the deflection angle is increasing for smaller values of $b$ but as $b$ increases the deflection angle become negatively decreasing.
\end{enumerate}
\end{itemize}
The photon lensing in weak field approximation has been studied
by many authors \cite{z5a}-\cite{ji}. Frolov \cite{R23} discussed the
collision of particles in the vicinity of horizon of weekly
magnetized non-rotating black hole in the presence of the
magnetic field Innermost Stable Circular Orbits (ISCO) of
charged particles. He demonstrated that for a collision
of two particles, one of which is charged and revolving at
ISCO and the other is neutral and falling from infinity,
the maximal collision energy can be high in the limit of
strong magnetic field. He also illustrated that for
 realistic astrophysical black holes, their ability to
play the role of accelerators is in fact quite restricted.
Liang \cite{J7} calculated the deflection angle in the strong
deflection limit and also obtained the angular positions
and magnifications of relativistic images as well as the
time delay between different relativistic images.
He also discussed the influence of the magnetic charge on
the black hole gravitational lensing. Recently, Turimov et al.
\cite{J8} have considered the magnetic field around a gravitational source
and illustrated that the split of the Einstein ring, as
the counterpart of the Zeeman effect. When the cyclotron
frequency approaches to the plasma frequency, the size and
the form of the ring change because of the presence of
a resonance state. This is a pure magnetic effect and can
potentially help to study magnetic fields through
gravitational lensing effects. Moreover, they found the
deflection angle of a photon moving in an inhomogeneous
magnetized plasma in the background of a static compact object
and the obtained deflection angle is
\begin{eqnarray*}
\alpha^{\pm}&\simeq& \frac{2M}{b}\left[1+(1-\frac{\omega_{0}^{2}}{\omega^{2}}-\frac{\omega_{0}^{2}}
{\omega^{2}}\frac{\omega_{c}}{\omega}f_{\pm}(\omega_{c},\omega_{0}))^{-1}\right]-
\frac{\omega_{0}^{2}}{\omega^{2}}\frac{\sqrt{\pi}\Gamma[(h+1)/2]}{\Gamma(h/2)}
\nonumber\\&\times&\left(\frac{R_{0}}{b}\right)^{h}\left(1-\frac{\omega_{0}^{2}}{\omega^{2}}-\frac{\omega_{0}^{2}}
{\omega^{2}}\frac{\omega_{c}}{\omega}f_{\pm}(\omega_{c},\omega_{0})\right)^{-1}+\mathcal{O}(M^{2}/b^{2})
\end{eqnarray*}
where $\Gamma(x)$ is the gamma function
\begin{equation*}
\Gamma(x)=\int_{0}^{\infty}t^{x-1}e^{-t}dt.
\end{equation*}

In our analysis, we study the weak gravitational
lensing and obtain the deflection angle of photon in the background of
magnetized black hole and analyze the effects of non-linear
electrodynamics by means of Gauss-Bonnet theorem.
To this end, initially, we set the photon rays on
the equatorial plane in the axisymmetric spacetime
and evaluate the corresponding optical metric.
After that we calculate the Gaussian optical
curvature for Gauss-Bonnet theorem. Furthermore,
we investigate the impact of correction
parameter, black hole charge and impact parameter
on deflection angle of photon for magnetized black
hole and investigate the effect of non-linear electrodynamics graphically.
We conclude that the mass $m$ decreases the deflection angle,
while the correction parameter $\beta$ increases the deflection angle.
Also, we prove that the impact parameter $b$ is directly proportional to the deflection angle.

In comparison with the deflection angle computed by Turimov et al. \cite{J8}, they found the
deflection angle of a light ray passing near a magnetized static compact object surrounded
by weak inhomogeneous plasma while we have evaluated the deflection angle by using GBT in a
weak gravitational lensing for spherically symmetric spacetime incorporating magnetic field
and correction parameter $\beta$. 
\section{Conclusion}
In this paper, we have analyze a model of NLED with parameter $\beta$. Then, we study the magnetized black hole and obtained the regular black hole solution. After that we calculate the optical Gaussian curvature for magnetized black hole. Then by using Gauss-Bonnet theorem, we calculate the weak gravitational lensing. We obtain the following angle of deflection for magnetized black hole
\begin{equation}
\Theta\thickapprox \frac{4mG}{b}-\frac{3\pi Gq^{2}}{4b^{2}}+\frac{7 G\pi \beta q^{4}}{32b^{6}}.
\end{equation}
We conclude that for $f(r)$ if $r\rightarrow\infty$ the space-time becomes flat, if $\beta=0$ the NLED model converted into Maxwell's electrodynamics and the solution becomes RN solution.
We have analyzed the behavior of deflection angle w.r.t impact parameter $b$, correction parameter $\beta$ and BH charge $q$.\\
The results obtained from the analysis of deflection angle given in the paper are summarized as follows:\\
\textit{\textbf{Deflection angle with respect to impact parameter:}}
\begin{itemize}
\item In our analysis we have to discussed the behavior of deflection angle, for this we choose different values of BH charge $q$ and fixed correction parameter. For smaller values, the deflection angle gradually decreasing but for large values of $q$ the deflection angle gradually decreases and then goes to infinity.
\item While for fixed $q$ and different values of $\beta$, the deflection angle constantly decreases for small range of $\beta$ and gradually decreases for greater values.
\item Thus, we conclude that for smaller values, BH indicates the stability and for large values shows the instability of BH.         
\end{itemize}
\textit{\textbf{Deflection angle with respect to correction parameter:}}
\begin{itemize}
\item The behavior of $\Theta$ w.r.t $\beta$, for fixed $q$ and varying $b$, positive behavior can be observed only for $b>0$ except $4\leq b<15$.
\item The behavior of $\Theta$ w.r.t $\beta$, for fixed $b$ and varying $q$, the behavior is positively increasing for $q>0$ except $5\leq q\leq10$.   
\end{itemize}
\textit{\textbf{Deflection angle with respect to black hole charge:}}
\begin{itemize}
\item The behavior of $\Theta$ w.r.t $q$, for fixed $b$ and varying $\beta$, the deflection angle initially decreases but with the increase of $\beta$, the deflection angle firstly decreases and then increases. For $\beta>20$ the deflection angle positively increases.
\item The behavior of $\Theta$ w.r.t $q$, for fixed $\beta$ and choose different values of $b$, the deflection angle positively increases for small $b$ and negatively decreases for greater $b$.      
\end{itemize}

\acknowledgments This work was supported by Comisi{\'o}n Nacional de Ciencias y Tecnolog{\'i}a of Chile through FONDECYT Grant $N^\mathrm{o}$
3170035 (A. {\"O}.).

\end{document}